\documentclass[prl,twocolumn,superscriptaddress]{revtex4}

\usepackage{amsmath,amsfonts,amssymb,amstext}
\usepackage{graphicx}
\usepackage{times}
\usepackage{bbm}
\usepackage{color}
\usepackage[colorlinks,linkcolor=blue]{hyperref}

\def\>{\rangle}
\def\<{\langle}
\def\({\left(}
\def\){\right)}

\newcommand{\ket}[1]{|#1\>}
\newcommand{\bra}[1]{\<#1|}

\renewcommand{\tensor}{\otimes}

\DeclareMathOperator{\Tr}{Tr}

\DeclareMathOperator{\poly}{poly}

\DeclareMathOperator{\rank}{rank}

\newcommand{\set}[1]{\lbrace #1 \rbrace}

\begin{document}

\title{Efficient Direct Tomography for Matrix Product States}

\author{Olivier Landon-Cardinal}
\affiliation{D\'epartement de Physique, Universit\'e de Sherbrooke, Sherbrooke, Qu\'ebec, Canada}

\author{Yi-Kai Liu}
\affiliation{Institute for Quantum Information, California Institute of Technology, Pasadena, CA, USA}

\author{David Poulin}
\affiliation{D\'epartement de Physique, Universit\'e de Sherbrooke, Sherbrooke, Qu\'ebec, Canada}

\date{February 23, 2010}

\begin{abstract}
In this note, we describe a method for reconstructing matrix product states
from a small number of efficiently-implementable measurements. Our method is
exponentially faster than standard tomography, and it can also be used to
certify that the unknown state is an MPS. The basic idea is to use local
unitary operations to measure in the Schmidt basis, giving direct access to
the MPS representation. This compares favorably with recently and independently proposed
methods that recover the MPS tensors by performing a variational minimization,
which is computationally intractable in certain cases. Our method also has the advantage of recovering any MPS, while other approaches were limited to special classes of states that exclude important examples such as GHZ and W states.
\end{abstract}

\maketitle


Matrix product states (MPS) \cite{AKLT87a,FNW92a,vidal03,Vid04a,Vid06a} are a variational class of states that can be specified by a small number of parameters. Their importance in quantum many-body physics and quantum information science stems from the fact that they seem to capture the low energy physics of a wide range of one-dimensional systems. Indeed, White's density matrix renormalization group \cite{Whi92a} can be seen as a variational method over this class of states \cite{OR95a,DMNS98a,VPC04b}.  

Quantum state tomography \cite{VR89a} is a procedure that uses many copies of a physical system  and performs a complete set of measurements on them in order to learn their quantum state. Because a generic quantum state of  a system comprising $n$ particles is described by a number of parameters growing exponentially with $n$, tomography requires exponentially many copies of the system on which to perform exponentially many distinct measurements. In addition, the outcome of these measurements must be processed on a computer in order to infer the state of the system; this post-processing also requires an exponential amount of time. While this method is clearly not scalable, it plays a central role in benchmarking quantum information processors and also in establishing a direct and complete link between theoretical models and physical systems.

Matrix product states, on the other hand, are described by a number of parameters polynomial (typically linear) in $n$. Thus, if a physical system is believed to be in an MPS---for instance a low temperature spin chain or a quantum simulator meant to prepare such a state---it should in principle be possible to learn its value from a small number of measurements. This would be a great advantage over standard quantum state tomography and would provide an essential tool for the experimental study of large scale MPS.

In this note, we present a tomography protocol to learn an MPS using a small number (polynomial in $n$) of efficiently implementable measurements. In addition, the post-processing required to infer the MPS from the measurement outcomes also scales polynomially in $n$. This contrasts with recent independent proposals for MPS tomography \cite{CP10a,FGBS10a} for which the post-processing needs to solve a problem which is in general intractable (NP-hard in some cases \cite{eisert06, schuch08}). Essentially, our method works by measuring the MPS tensors \textit{directly}, while the other methods work by estimating the local density matrices of the state, then performing a constrained variational minimization over all MPS states. 

Our method has the additional advantage of recovering any MPS, while the other approaches were limited to special classes of states (injective MPS \cite{PVWC07a}) that exclude important examples such as GHZ and W states. As a drawback, our method requires quantum circuits of size polynomial in $n$, while the methods of \cite{CP10a,FGBS10a} use quantum circuits whose size is logarithmic in the bond dimension $\chi$. (Note, however, that the modification proposed in \cite{FGBS10a} to handle GHZ and W states does require quantum circuits of size $\poly(n)$.)


\paragraph{Direct MPS tomography}
In what follows, we describe our method for an open-boundary chain of $n$ particles, where each particle is a $d$-dimensional qudit. Generalization to the case of a periodic boundary ring is straightforward (with an increased measurement and post-processing complexity).   Let $\ket{\psi}$ be the matrix product state of this system, with bond dimension $\chi$.  We are given many copies of $\ket{\psi}$, and our goal is to determine its MPS representation.  

The key fact is that $\ket{\psi}$ has Schmidt rank at most $\chi$, i.e,  for any partition of the chain into two segments $\set{1,\ldots,l}$ and $\set{l+1,\ldots,n}$, the state can be written as
\begin{equation}
\ket{\psi} = \sum_{j=1}^\chi \alpha_j \ket{\phi^{1,\ldots, l}_j}\otimes \ket{\phi^{l+1,\ldots n}_j}
\end{equation}
where $\ket{\phi^{1,\ldots, l}_j}$ are orthogonal states of the $l$ first particles and $\ket{\phi^{l+1,\ldots n}_j}$ are orthogonal states on the last $n-l$ particles.  Now, look at the first $k$ sites, choosing $k$ such that $d^{k-1} \geq \chi$.  The reduced density matrix on these sites has rank at most $\chi$, so by performing a unitary operation, we can rotate the state into the subspace $\ket{0} \tensor (\mathbb{C}^d)^{\tensor k-1}$.  This disentangles the first qudit from the rest of the chain.  Now we set aside the first qudit, look at sites $2,\ldots,k+1$, and recurse.  

In this way, we get a sequence of unitary operations $U_1, U_2, \ldots U_{n-k+1}$, where $U_i$ acts on sites $i,\ldots,i+k-1$.  This sequence of operations transforms $\ket{\psi}$ into the state $\ket{0}^{\tensor n-k+1} \ket{\eta}$, where $\ket{\eta}$ is some pure state on the last $k-1$ sites.  

\vspace{10pt}
\ifx\JPicScale\undefined\def\JPicScale{1}\fi
\unitlength \JPicScale mm
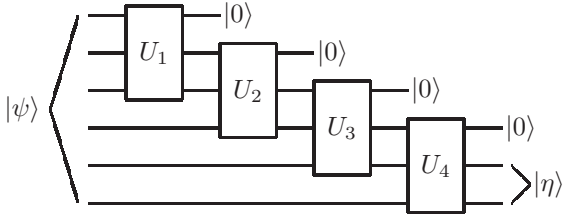
\begin{picture}(75,31.25)(0,0)
\put(13.75,0){\makebox(0,0)[cc]{}}

\linethickness{0.3mm}
\put(13.75,30){\line(1,0){5}}
\linethickness{0.3mm}
\put(13.75,25){\line(1,0){5}}
\linethickness{0.3mm}
\put(13.75,20){\line(1,0){5}}
\linethickness{0.3mm}
\put(18.75,31.25){\line(1,0){7.5}}
\put(18.75,18.75){\line(0,1){12.5}}
\put(26.25,18.75){\line(0,1){12.5}}
\put(18.75,18.75){\line(1,0){7.5}}
\put(22.5,25){\makebox(0,0)[cc]{$U_1$}}

\linethickness{0.3mm}
\put(26.25,25){\line(1,0){5}}
\linethickness{0.3mm}
\put(26.25,20){\line(1,0){5}}
\linethickness{0.3mm}
\put(13.75,15){\line(1,0){17.5}}
\linethickness{0.3mm}
\put(31.25,26.25){\line(1,0){7.5}}
\put(31.25,13.75){\line(0,1){12.5}}
\put(38.75,13.75){\line(0,1){12.5}}
\put(31.25,13.75){\line(1,0){7.5}}
\put(35,20){\makebox(0,0)[cc]{$U_2$}}

\linethickness{0.3mm}
\put(38.75,20){\line(1,0){5}}
\linethickness{0.3mm}
\put(38.75,15){\line(1,0){5}}
\linethickness{0.3mm}
\put(13.75,10){\line(1,0){30}}
\linethickness{0.3mm}
\put(43.75,21.25){\line(1,0){7.5}}
\put(43.75,8.75){\line(0,1){12.5}}
\put(51.25,8.75){\line(0,1){12.5}}
\put(43.75,8.75){\line(1,0){7.5}}
\put(47.5,15){\makebox(0,0)[cc]{$U_3$}}

\linethickness{0.3mm}
\put(51.25,15){\line(1,0){5}}
\linethickness{0.3mm}
\put(51.25,10){\line(1,0){5}}
\linethickness{0.3mm}
\put(13.75,5){\line(1,0){42.5}}
\linethickness{0.3mm}
\put(56.25,16.25){\line(1,0){7.5}}
\put(56.25,3.75){\line(0,1){12.5}}
\put(63.75,3.75){\line(0,1){12.5}}
\put(56.25,3.75){\line(1,0){7.5}}
\put(60,10){\makebox(0,0)[cc]{$U_4$}}

\linethickness{0.3mm}
\put(26.25,30){\line(1,0){5}}
\put(33.75,30){\makebox(0,0)[cc]{$\ket{0}$}}

\linethickness{0.3mm}
\put(38.75,25){\line(1,0){5}}
\put(46.25,25){\makebox(0,0)[cc]{$\ket{0}$}}

\linethickness{0.3mm}
\put(51.25,20){\line(1,0){5}}
\put(58.75,20){\makebox(0,0)[cc]{$\ket{0}$}}

\linethickness{0.3mm}
\put(63.75,15){\line(1,0){5}}
\put(71.25,15){\makebox(0,0)[cc]{$\ket{0}$}}

\linethickness{0.3mm}
\put(63.75,10){\line(1,0){5}}
\put(75,7.5){\makebox(0,0)[cc]{$\ket{\eta}$}}

\linethickness{0.3mm}
\put(63.75,5){\line(1,0){5}}
\linethickness{0.3mm}
\multiput(70,10)(0.12,-0.12){21}{\line(1,0){0.12}}
\linethickness{0.3mm}
\multiput(70,5)(0.12,0.12){21}{\line(1,0){0.12}}
\linethickness{0.3mm}
\multiput(8.75,17.5)(0.12,0.4){31}{\line(0,1){0.4}}
\linethickness{0.3mm}
\multiput(8.75,17.5)(0.12,-0.4){31}{\line(0,-1){0.4}}
\put(5,17.5){\makebox(0,0)[cc]{$\ket{\psi}$}}

\put(5,18.75){\makebox(0,0)[cc]{}}

\end{picture}

\noindent
Note that this is precisely the procedure for preparing an MPS state, running in reverse \cite{schoen05}.  From this decomposition of $\ket{\psi}$, 
\begin{equation}
\ket{\psi} = U_1^{-1} \cdots U_{n-k+1}^{-1} \ket{0}^{\tensor n-k+1} \ket{\eta},
\end{equation}
one can write down the MPS tensors.  We have 
\begin{equation} \label{eqn-mps-1}
\ket{\psi} = \sum_{z_1,\ldots,z_n \in \set{0,\ldots,d-1}} c_{z_1,\ldots,z_n} \ket{z_1,\ldots,z_n}, 
\end{equation}
where 
\begin{align} \label{eqn-mps-2}
c_{z_1,\ldots,z_n} 
 &= \bra{z_1,\ldots,z_n} U_1^{-1} \cdots U_{n-k+1}^{-1} \ket{0}^{\tensor n-k+1} \ket{\eta}.  
\end{align}
As we will show below, this can be written as a product of operators acting on $(\mathbb{C}^d)^{\tensor k-1}$.

In detail, our method is as follows.  Let $k = \lceil \log_d\chi \rceil + 1$.  For $i=1,\ldots,n-k+1$, do the following steps:
\begin{enumerate}
\item Prepare the state $\ket{\psi'} = U_{i-1} \cdots U_1 \ket{\psi}$.

\item Perform standard tomography on sites $i$ through $i+k-1$, to recover the reduced density matrix $\rho' = \Tr_{1,\ldots,i-1; i+k,\ldots,n} \ket{\psi'}\bra{\psi'}$.

\item Let $\ket{\phi'_j}$ ($j=1,\ldots,\rank(\rho')$) be the eigenvectors of $\rho'$.  Note that $\rank(\rho') \leq \chi \leq d^{k-1}$.  Define additional vectors $\ket{\phi'_j}$ for $j=\rank(\rho')+1,\ldots,d^k$, such that the $\ket{\phi'_j}$ form an orthonormal basis for $(\mathbb{C}^d)^{\tensor k}$.

\item Define the unitary 
\begin{equation}
U_i = \sum_{a=0}^{d-1} \sum_{j=1}^{d^{k-1}} \ket{a}\ket{j}\bra{\phi'_{ad^{k-1}+j}}, 
\end{equation}
acting on sites $i$ through $i+k-1$.  
\end{enumerate}

We can reconstruct the MPS representation of $\ket{\psi}$ as follows:  For $i=1,\ldots,k-1$, and $z\in\set{0,\ldots,d-1}$, define $T_i^z$ to be an operator on $(\mathbb{C}^d)^{\tensor k-1}$, which is given by 
\begin{equation}
T_i^z = I_{1,\ldots,i-1} \tensor \Bigl(\ket{0}\bra{z}\Bigr)_i \tensor I_{i+1,\ldots,k-1}.  
\end{equation}
$T_i^z$ acts nontrivially on the $i$'th qudit, and acts as the identity on the other qudits.  

For $i=1,\ldots,n-k+1$, and $z\in\set{0,\ldots,d-1}$, we define $V_i^z$ to be an operator on $(\mathbb{C}^d)^{\tensor k-1}$, which is given by 
\begin{equation}
V_i^z = \Bigl( I_{1,\ldots,k-1} \tensor \bra{z}_k \Bigr) U_i^{-1} \Bigl( \ket{0}_1 \tensor I_{2,\ldots,k} \Bigr).  
\end{equation}
Here we view $U_i^{-1}$ as an operator on $(\mathbb{C}^d)^{\tensor k}$, and on the left we apply $\bra{z}$ on the last qudit, and on the right we apply $\ket{0}$ on the first qudit.

Then equation (\ref{eqn-mps-2}) can be rewritten as:  
\begin{equation}
c_{z_1,\ldots,z_n} = \bra{0}^{\tensor k-1} T_1^{z_1} \cdots T_{k-1}^{z_{k-1}} V^{z_k}_1 \cdots V^{z_n}_{n-k+1} \ket{\eta}.
\end{equation}
So $T_i^z$ and $V_i^z$ are the desired MPS tensors, with boundary conditions given by $\bra{0}^{\tensor k-1}$ and $\ket{\eta}$.  Note that the bond dimension is $d^{k-1}$, which is at most $d\cdot\chi$.  It is easy to check that this procedure uses $\poly(n,\chi)$ local unitary gates, and the classical postprocessing takes time $\poly(n,\chi)$.

\paragraph{Accuracy}
There are a few factors that limit the accuracy of MPS tomography in an actual experiment.  In particular, the state of the system is generally not exactly an MPS, but merely close to one; and the measurements are noisy and restricted to finite precision.  We will address each of these issues in turn.  

First, we slightly modify our tomography procedure.  Note that the reduced density matrix on sites $1,\ldots,l$ will actually be full-rank, though most of its probability mass will lie on a subspace of dimension at most $\chi$.  So, each time we apply a unitary $U_i$, we also want to \textit{truncate} the reduced state on sites $i,\ldots,i+k-1$ to a subspace of dimension $d^{k-1}$.  A simple way is to measure site $i$ in the standard basis, and postselect on the $\ket{0}$ outcome (which occurs with high probability).  Note that we know explicitly how large is this truncation error, since we know the reduced density matrix on sites $i,\ldots,i+k-1$.

Clearly, if $\ket{\psi}$ is close to a matrix product state, the truncation errors will be small.  More importantly, if $\ket{\psi}$ is far from any MPS, the truncation errors will be large.  Thus, we can do certified tomography, i.e., we can test whether an arbitrary state $\ket{\psi}$ is close to an MPS.

Now consider the effect of noisy measurements and finite precision.  (Suppose, for simplicity, that the state $\ket{\psi}$ is exactly an MPS.)  Our estimate of the density matrix $\rho'$ on sites $i,\ldots,i+k-1$ will be slightly wrong, and hence our estimate of the unitary $U_i$ will be slightly wrong.  In addition, the error in $U_i$ can result in a truncation error when we measure site $i$ and postselect on the $\ket{0}$ outcome.  This truncation error affects the subsequent iterations of our procedure, where we estimate $U_{i+1}$, $U_{i+2}$ and so on.  

More concretely, let $U_i$ be the unitary matrices for the true state $\ket{\psi}$ and let $\tilde U_i$ be the estimates produced by our procedure. Note that in order to represent the same MPS, these matrices need only to coincide up to some gauge transformation, which can be fixed arbitrarily. Equivalently, note that while it is crucial to obtain a good estimate of the $d^{k-1}$-dimensional subspace on which $\rho'$ is supported, it is less important to identify the individual eigenvectors of $\rho'$ exactly. Assume that we estimate this subspace within error $\epsilon$. Then, fixing the gauge, we have $\|U_i-\tilde U_i\| \leq i\cdot\epsilon$ (where the linear dependence on $i$ comes from the accumulation of truncation errors from previous iterations). 
This implies that the reconstructed MPS $\ket{\tilde\psi}$ will be close to the true state of the system, 
\begin{equation}
\bigl\lVert \ket{\psi}-\ket{\tilde\psi} \bigr\rVert \leq n^2\epsilon. 
\end{equation}

\paragraph{Local measurements degeneracy}
The schemes for MPS tomography proposed in \cite{CP10a,FGBS10a} make use of local measurements on small intervals of the chain. The outcome of these measurements impose constraints on the global MPS.  For instance, if one chooses these intervals to be of size $2k$, the reduced density matrix on any such interval has non-maximal rank due to the MPS structure.  A minimization procedure is then employed to find the MPS that satisfied all these constraints.

We observe that this procedure will fail in cases where the MPS is not injective. This situation will occur for instance when the system is in the ground state of a degenerate Hamiltonian. In those cases, the information gathered by any local measurement is not sufficient to recover the global state: the solution to the minimization problem is not unique, it is degenerate. A simple example is given by GHZ states of $n$ qubits, i.e. $\ket\psi = a \ket0^{\otimes n} + e^{i\varphi} b \ket 1^{\otimes n}$. Local measurements are insensitive to the relative phase $\varphi$, so the methods of \cite{CP10a,FGBS10a} fail to reconstruct this state. On the other hand, the final "boundary" state obtained at the $n$th step of our protocol will be $\ket\eta = a\ket 0 + e^{i\varphi}\ket 1$, so the phase information is revealed during the final step of tomography. The same situation would occur with $W$ states $\ket\psi = \frac1{\sqrt n} \left( e^{i\varphi_1}\ket{100\ldots0} + e^{i\varphi_2}\ket{010\ldots0} + \ldots e^{i\varphi_n}\ket{000\ldots1}\right)$.

In \cite{FGBS10a}, a fix was proposed to overcome this problem under the assumption that the degeneracy is independent of $n$. Their method is to identify the entire degenerate subspace and to perform further measurements inside this subspace to identify the state. We note that convolutional quantum error correcting codes \cite{Cha98b,OT03a} are an important class of exponentially degenerate MPS for which this approach would fail. Our method is suitable for this particular situation as well as any other MPS. In the case of convolutional codes, the unitaries $U_i$ returned by our procedure would coincide with the encoding circuit of the code.

\paragraph{Conclusion}
We have presented a very simple, direct method to perform quantum state tomography of matrix product states using a small number of measurements. In our scheme, there is no need to perform numerical optimization to recover the MPS from experimental data; this is its main advantage over recently proposed methods. It also has the advantage of recovering any MPS, and is not limited to non- or constant-degenerate cases. Like other approaches, our scheme can be used to certify that the unknown state is indeed close to an MPS with given bond dimension. Finally, our scheme extends trivially to the case of mixed states for which the reduced state on any interval has bounded rank.

\paragraph{Acknowledgements} 
We thank Steve Flammia, Steve Bartlett, and David Gross for enjoyable discussions. OLC is partially funded by FQRNT. DP is partially funded by NSERC and FQRNT.



\end{document}